# An Efficient Large-Area Grating Coupler for Surface Plasmon Polaritons

*Stephan T. Koev[1,2], Amit Agrawal[1,2,3], Henri J. Lezec[1], and Vladimir Aksyuk[1,*]*

[1]*Center for Nanoscale Science and Technology, National Institute of Standards and Technology, Gaithersburg, MD 20899, USA*

[2]*Maryland Nanocenter, University of Maryland, College Park, MD 20742, USA*

[3]*Department of Electrical Engineering and Computer Science, Syracuse University, Syracuse, NY 13244, USA*

*Corresponding Author: vaksyuk@nist.gov

**Abstract:** We report the design, fabrication and characterization of a periodic grating of shallow rectangular grooves in a metallic film with the goal of maximizing the coupling efficiency of an extended plane wave (PW) of visible or near-infrared light into a single surface plasmon polariton (SPP) mode on a flat metal surface. A PW-to-SPP power conversion factor > 45 % is demonstrated at a wavelength of 780 nm, which exceeds by an order of magnitude the experimental performance of SPP grating couplers reported to date at any wavelength. Conversion efficiency is maximized by matching the dissipative SPP losses along the grating surface to the local coupling strength. This critical coupling condition is experimentally achieved by tailoring the groove depth and width using a focused ion beam.

## Introduction

The field of nanoplasmonics holds tremendous potential for novel optical devices exploiting the confinement of light to subwavelength dimensions [1-3]. In particular, during the last decade there has been overwhelming interest in using devices based on surface plasmon polaritons (SPPs). SPPs are collective charge oscillations coupled to an external electromagnetic field that propagate along an interface between a metal and a dielectric [4]. Despite the rapid progress in nanoplasmonics, a number of challenges remain that hinder the development of practical applications. One of these challenges is efficient coupling between free-space light and SPPs on a flat metal surface [5]. This coupling is commonly performed with a diffraction grating, which shifts the wave vector of an incident plane wave to match the SPP wave vector [1, 6-8]. Although other techniques such as prism coupling [9], nearfield excitation [10], and end-fire coupling [11] exist, the grating coupler approach is attractive for device integration since the grating can be fabricated directly on the metal film supporting the SPPs [12].

There has been a wealth of theoretical work describing the operation of grating couplers and approaches to improve their efficiency [13-20]. However, these reports generally have not been accompanied with experimental studies that validate the theories. Several groups have experimentally demonstrated efficient light-to-SPP coupling by using a tightly focused beam (instead of a plane wave) incident on isolated scattering structures (instead of periodic gratings) [21-23]. However, this strategy produces a point source of omnidirectional SPPs. In contrast, using a grating coupler and an incident plane wave (PW) results in a collimated, unidirectional SPP beam [1, 6, 24] that can propagate with low divergence to other devices on the same chip. Furthermore, PW illumination has higher translational alignment tolerance than focused illumination, which simplifies the experimental setup and enables low-cost packaging of plasmonic devices.

The theoretically proposed grating couplers mentioned above reach very high efficiencies (> 60 % some cases) [13]. However they require use of elaborate geometries, such as slanted [13], variable-width [14], or sinusoidal [15] grooves, that are difficult to fabricate. In this paper, we present an analytical model for the optimization of a simple "classical" grating coupler (identical rectangular grooves on an Au film), and demonstrate experimentally a device with PW-to-SPP coupling efficiency > 45 % at 780 nm and > 30 % at 675 nm. These values are an order of magnitude higher than the highest published experimental coupling efficiency to date for a grating coupler [8] (3.5 % at 675 nm). We achieve this significant improvement in efficiency by appropriate choice of the groove spacing and dimensions. Specifically, the grating period and incidence angle are chosen to minimize power loss into multiple diffraction orders. The groove depth and width are chosen to achieve a critical coupling condition, whereby reflection from the grating is suppressed and maximum power is transferred to the SPPs. We fabricate an array of devices where the groove width and depth are varied independently in order to study this new coupling regime experimentally and converge on the best coupling efficiency. Our simple theoretical model explaining the experimental results presented here may serve as a useful guide for creating practical, high-efficiency grating couplers for SPPs.

## Design

The grating coupler used in this study consists of an array of parallel grooves on an Au film as illustrated in Fig. 1a. The grating can serve as an input coupler, which couples a PW mode into a SPP mode on the grating and subsequently into an SPP mode on the flat film. Alternatively, it can serve as an output coupler, which couples the SPP mode on the flat film to multiple PW modes. We focus on the optimization of the input coupling mechanism only; the output coupling depends on the acceptance angle of the collection optics used and is therefore highly application dependent. The input coupling efficiency is defined here as the ratio of SPP power in the flat film to the power of free-space light incident on the grating. The output efficiency is defined as the ratio of optical power entering the detector to the SPP power entering the grating.

An in-depth treatment of grating coupler theory is beyond the scope of this paper and can be found elsewhere [1, 4, 16, 25]. In the following, we start with simple theoretical guidelines for the selection of groove spacing, and later we present a more thorough model that guides the selection of groove size. The PW incident on the grating is scattered at each groove and locally excites SPPs. At the correct incidence angle and groove spacing, these excitations interfere constructively, and maximum coupling to the SPP mode on the grating is achieved [1]. This coupling condition is summarized by Eq. 1 for incident light with wavelength $\lambda$ in vacuum, where $k_0 = 2\pi/\lambda$ and $k_{SPPgr}$ are the magnitudes of the free-space wave vector and grating SPP wave vector respectively, $\theta_i$ is incidence angle, $d$ is the grating period, and $n$ is a positive or negative integer.

$$k_{SPPgr} = -k_0 \sin\theta_i + 2\pi n/d \quad (1)$$

To enable PW to SPP coupling for a given incident wavelength, $d$ is chosen such that there is at least one angle satisfying Eq. 1, which leads to $\lambda/d < k_{SPPgr}/k_0 + 1$. A rigorous solution to this problem would be difficult since $k_{SPPgr}$ has an elaborate dependence on grating geometric parameters (including $d$). However, in the case of shallow gratings, $k_{SPPgr}$ is close to $k_{SPPff}$ (the SPP wave vector on the flat film) and does not vary much with grating geometry. For example, we observe experimentally that $k_{SPPgr}$ changes by less than 3 % when changing grating depth from 30 nm to 80 nm ($k_{SPPgr}$ is calculated from measured values of $\theta_i$). Therefore, Eq. 1 can be used for design purposes by approximating $k_{SPPgr}$ with $k_{SPPff}$. Using typical Au permittivity values reported in the literature [26], we calculate $k_{SPPff} = 1.02\ k_0$ at 780 nm wavelength and $k_{SPPff} = 1.04\ k_0$ at 675 nm.

Not all solutions to Eq. 1 are favorable for high-efficiency coupling. Surface-normal incidence should be avoided as, by symmetry, it splits the power equally into forward and backward propagating SPPs. Additionally, the incidence angle that excites the SPPs must be such that except specular reflection, there are no diffraction orders coupling power into free space as they would constitute a channel for power loss. Eq. 2 shows the condition for diffraction into free space, where $\theta_m$ is the angle of the $m^{th}$ diffraction order.

$$k_0 \left( \sin \theta_m - \sin \theta_i \right) = 2\pi m / d \quad (2)$$

By proper choice of $d$ and $\theta_i$, we can force this equation to have no solution for all $m \neq 0$ while still satisfying the SPP coupling condition given by Eq. 1. We choose $n = 1$ and $\theta_i > 0$ such that, from Eq. 1, $2\pi/d > k_{SPPgr}$. Given that $k_{SPPgr} > k_0$, this leads to $\lambda/d > k_{SPPgr} / k_0 > 1$. Under these conditions, Eq. 2 has no real solution for $\theta_m$ with non-zero $m$, i.e. there is only specular reflection into free space with $\theta_0 = \theta_i$. As explained previously, Eq. 1 only has a real solution for $\theta_i$ if $\lambda/d < k_{SPPgr}/k_0 + 1$. Accordingly, the range of favorable groove spacing for a grating coupler can be summarized by the relation $k_{SPPgr}/k_0 < \lambda/d < k_{SPPgr}/k_0 + 1$. Note that this range is quite broad, and we can choose a proper $d$ even by using an approximate value for $k_{SPPgr}$.

In the following, we present a more thorough model of the grating coupler that helps us to optimize the groove size for peak input coupling efficiency. Consider the amplitudes of the electromagnetic modes interacting with a single groove as shown in Fig. 1b (the SPP modes denoted here are modes on the flat metal film between grooves; the grating SPP mode referenced previously is a superposition of these modes). The stronger the scattering by the groove is, the more free-space light is scattered into the SPP; however, this also means that a larger fraction of the SPP mode is scattered out upon arrival at a neighboring groove. The following analysis addresses this tradeoff and gives a prescription for optimal scattering strength. We assume that the groove scattering increases monotonically with groove width and depth, i.e. there are no resonant peaks in the scattering as a function of groove dimensions.

Let the grating be uniformly illuminated with a plane wave with electric field amplitude $a_0$. If the dissipative loss at each groove is small, the incident and scattered electric fields for the $m^{th}$ groove can be related with a unitary scattering matrix as shown in Eq. 3 ($a_{1,2}$ and $b_{1,2}$ are defined in Fig. 1b; $\tau$ and $\kappa$ are scattering coefficients). The SPP propagation between adjacent grooves is modeled with Eq. 4, where $\theta$ and $\alpha$ are the phase shift and dissipative loss terms, respectively. The free-space light incident on each groove has the same amplitude and a varying phase given by Eq. 5, where $\varphi$ is the phase shift of this light between adjacent grooves.

$$\begin{bmatrix} b_1^m \\ b_2^m \end{bmatrix} = \begin{bmatrix} -\tau & \kappa \\ \kappa^* & \tau^* \end{bmatrix} \begin{bmatrix} a_1^m \\ a_2^m \end{bmatrix} \quad (3)$$

$$a_2^{m+1} = b_2^m \alpha \exp(i\theta) \quad (4)$$

$$a_1^m = a_0 \exp(-im\varphi) \quad (5)$$

For a shallow grating the SPP power change per grating period is small and we can approximate a periodic grating by a continuum. Substituting the derivative $da_2/dx$ for $(a_2^{m+1} - a_2^m)/d$, we obtain Eq. 6. Here $x = md$, $L_a = -d\ln(\alpha)$, $L_s = -d\ln(|\tau|)$, $k_v = \theta/d$, $k_\tau = \arg(\tau)/d$, and $k_a = \varphi/d$. Solving Eq. 6 under phase-matched condition ($k_v - k_\tau = k_a$) equivalent to Eq. 1 results in Eq. 7 and Eq. 8, where $A_0^2 = (a_0)^2/d$ and $B^2 = (b_1)^2/d$ are intensities of light incident on the grating and reflected by the grating, respectively.

$$\frac{da_2}{dx} = \frac{a_0}{\sqrt{dL_s/2}} \exp(-ik_a x) - a_2 \left( \frac{1}{L_s} + \frac{1}{L_a} \right) + ia_2 (k_v - k_\tau) \quad (6)$$

$$|a_2(x)|^2 = \frac{2A_0^2}{L_s (1/L_a + 1/L_s)^2} \left( 1 - e^{-x(1/L_a + 1/L_s)} \right)^2 \quad (7)$$

$$|B(x)|^2 = A_0^2 \left( \frac{2}{(L_s/L_a + 1)} \left( 1 - e^{-x(1/L_a + 1/L_s)} \right) - 1 \right)^2 \quad (8)$$

Eq. 7 represents the SPP power. It builds up from 0 at the beginning of the grating to a steady state for large $x$. The steady state power in Eq. 7 is a function of $L_a$ and $L_s$; these are SPP amplitude decay lengths inversely related to the absorption loss and the groove scattering strength, respectively. In practice, $L_a$ is determined mainly by material properties and there is little control over it. $L_s$ is controlled by geometric parameters (e.g. groove width and depth) and can be varied during fabrication. To maximize the steady state SPP power for a fixed $L_a$, we set its derivative to 0, which yields $L_s = L_a$. This critical coupling condition corresponds to the scattering strength needed for maximum coupling efficiency; it is equivalent to stating that the radiative SPP propagation loss in the grating should equal the nonradiative absorption loss. This condition is analogous to the maximum power transfer theorem in electric circuits, which states that a voltage source with a fixed source resistance transfers maximum power to a load when the load resistance equals the source resistance.

Eq. 8 represents the reflected light intensity from the grating. It is the result of interference between specular reflection and light scattered out by the propagating SPPs. Plotting $|B(x)|^2$ gives the profiles in Fig. 2. If $L_s < L_a$ (strongly scattering), the intensity first goes down and then builds up to a steady-state value. If $L_s > L_a$ (weakly scattering), the intensity goes monotonically down to a steady-state value. When $L_s = L_a$, the steady state reflected power is asymptotically smallest due to the destructive interference between the out-coupled SPP and the specular reflection. Therefore, by plotting the measured intensity from a microscope image of the input grating under illumination, one can determine whether the grooves are too large or too small for optimal coupling. This is useful because translating the equality $L_s = L_a$ into an actual grating design may not be feasible analytically. Both the scattering strength and SPP propagation loss are influenced by fabrication artifacts such as line edge roughness and surface roughness, and they are non-trivial to model accurately. By fabricating several test gratings with varying groove widths and depths, one can determine the optimal groove dimensions. In the following, we experimentally demonstrate the validity of this approach.

The analysis presented here assumes that the grating is illuminated with a single PW mode, which is experimentally approximated by a Gaussian beam with waist larger than the grating dimensions. A strongly diverging beam contains multiple PW modes; only the mode that satisfies the coupling condition (Eq. 1) is appreciably coupled into SPPs while the others are reflected, thereby decreasing the grating efficiency. In addition, the reflected modes change the expected intensity profile in Fig. 2. Therefore, in order to use this profile as an indicator of scattering strength, it is essential to have a well collimated beam incident on the grating.

The optimal efficiency condition derived here is based on maximizing the steady-state SPP power given by Eq. 7. This approach is valid when the grating is considerably longer than $L_a$ and is completely illuminated. A closer look at the equation suggests that we can achieve higher efficiency by increasing $A_0$ and reducing grating length. This is equivalent to reducing the incident beam diameter to achieve higher intensity and using a shorter grating which ends before the SPP has reached its steady state power. However, this approach is dependent on the experimental setup (e.g. incident beam divergence as a function of diameter) and introduces other tradeoffs (e.g. reduction in coupling efficiency due to beam divergence). Motivated to obtain efficient coupling to collimated SPP from large area PW, in this work we use a large-diameter incident beam and gratings sufficiently long to reach steady state SPP power.

## Fabrication

To demonstrate the optimization approach described above, we fabricated and tested grating couplers with varying groove scattering strengths. Measurements were performed at wavelengths of 780 nm (results presented in this paper) and 675 nm (presented as electronic supplementary material). The gratings consist of identical parallel grooves of varying width and depth milled by a focused ion beam (FIB) into a 300 nm thick atomically smooth Au film on a mica substrate (Fig. 3a). The etch depth was controlled by varying the FIB ion dose; the depth as a function of dose was estimated by milling 5 μm wide test features and measuring their depth with an optical profiler (the measurement uncertainty of that instrument based on two standard deviations is less than ± 2 nm).

The gratings were grouped in input-output pairs as shown in Fig. 3b. Following the analysis described earlier, the groove spacing was chosen to be 560 nm to suppress diffraction into air. All output gratings were identical, with groove width of 110 nm and groove depth of 50 nm. Each input grating had a fixed groove width and depth, which were independently varied from grating to grating. All combinations of three different groove widths (60 nm, 110 nm, 190 nm) and 7 different groove depths (28 nm, 38 nm, 47 nm, 56 nm, 69 nm, 79 nm, 117 nm) were explored, yielding 21 different input gratings (and the same number of grating pairs). This configuration is designed to keep the output efficiency constant thus allowing direct measurement of changes in the input efficiency as the cross-sectional geometries of the grooves of the input gratings are varied. In addition, another sample with 5 pairs of identical input and output gratings was fabricated, where the edge-to-edge spacing between the input and output gratings was varied to take on values of 40, 55, 70, 85, and 100 μm. This configuration is designed to maintain constant efficiencies for the input and output couplers, thus enabling direct estimation of the SPP propagation decay length in the flat Au film between the gratings.

## Experimental Setup

The experimental setup used to test the grating couplers is based on an inverted optical microscope with a 40x objective (NA = 0.95) and high-resolution Si-CCD camera (Fig. 4). Several custom modifications were made to the microscope to enable illumination of the input grating with a laser beam. A beamsplitter was introduced between the objective and the tube lens; the laser beam was focused through this beam splitter onto the back focal plane of the objective. This produces a collimated beam in the sample plane. Before entering the beam splitter, the laser beam was reflected off a tilt mirror positioned at a conjugate focal plane and passed through a variable aperture (with diameter much smaller than the beam diameter) and a polarizer. Adjusting the mirror changes the incidence angle of the beam on the sample without moving the beam spot appreciably. Adjusting the aperture changes the spot size on the sample. The polarizer blocks the s-polarization, which cannot excite SPPs in our configuration and is therefore not useful. The described setup produces a round area of illumination on the sample surface that completely covers the input grating while the output grating remains outside. By using only the center portion of the strongly expanded Gaussian beam and focusing it at the objective back focal plane, the illumination is designed to have uniform intensity and a planar wavefront over the grating area, approximating a PW.

Fig. 3c displays a microscope image of an input-output grating pair under illumination of the input grating at a wavelength of 780 nm with the incidence angle adjusted for maximum SPP coupling. From similar images, we obtain the two-dimensional intensity profiles over the surface of the input and output gratings, as well as the integrated power emitted into free space from the output grating. For normalization, we also obtain the power incident on the input grating by moving the laser spot to an unpatterned Au area, integrating the reflected intensity over the fraction of the illuminated surface area previously occupied by the input grating, and finally equating the reflected power to the incident power (assuming Au film reflectivity $\approx$ 100 % at both explored wavelengths [26]). Dividing the measured output power by the measured input power gives the power transmission coefficient of the grating coupler pair. This coefficient is equal to the product of the efficiencies of the two couplers and the SPP power transmission coefficient of the flat Au film between the couplers (the latter is the ratio of SPP power entering the output grating to the SPP power emitted from the input grating into the Au film).

## Results and Discussion

First, we measured the SPP power decay length on the flat Au film between the gratings. This decay length is dominated by the nonradiative absorption loss in the Au. Using the sample with varied spacing between the input and output gratings, we measured the transmission coefficient of the grating pair vs. spacing. A fit of the data with a decaying exponential yields a 1/e SPP power decay length of 62.5 ± 1.8 μm at 780 nm and 21.3 ± 0.5 μm at 675 nm (with an uncertainty based on 2 standard deviations of 10 measurements). These measured values are larger than the decay lengths calculated using conventional formulas for SPP dispersion [4] and tabulated Au dielectric properties [26], which are 39.5 μm and 13.4 μm at the two respective wavelengths. This discrepancy is most likely due to the difference in the dielectric properties between the conventionally deposited polycrystalline Au films on glass used in the calculation and the atomically smooth near single crystal Au films on mica used in our experiments.

Next, we measured the SPP power decay length $L_d$ on the 21 different input grating surfaces. This length, equal to the term $(1/L_a+1/L_s)^{-1}/2$ in Eq. 8, depends on radiative loss (due to scattering by the grooves) in addition to the nonradiative absorption loss in the Au and thus is expected to be substantially shorter than the decay length measured above for propagation on a flat Au film. $L_d$ was extracted from the measured image intensity of a grating as follows. The intensity profile of an output grating is $exp(-2 \cdot x \cdot (1/L_a+1/L_s))$; this profile is similar to that of an input grating (Eq. 8, Fig. 2) but is simpler due to the absence of reflected light at the output. To find the SPP decay length for each of the different input gratings, we tested the grating pairs in reverse, i.e. used the side with variably sized grooves as the output while keeping the input coupler geometry and hence efficiency constant. We plotted the output intensity profile, fitted it with an exponential curve, and extracted the 1/e decay length. The results at 780 nm are shown in Fig. 5a. The error bars represent the range of decay lengths (maximum and minimum) obtained by fitting different regions within the intensity profile (it is not perfectly exponential due to laser speckle).

The trends observed in Fig. 5a are intuitively expected. Increasing the groove width or depth increases the SPP scattering, thereby reducing the decay length. It is possible to achieve a certain decay length with multiple combinations of width and depth.

Next, we compared the coupling efficiencies of the 21 different grating pairs. We measured the transmission coefficient of the coupler pair as previously described (variably sized grooves on the input side) and divided it by the transmission coefficient of the flat Au film between the gratings (calculated from the measurement of SPP decay length on the flat Au film). The result is the product of input and output coupling efficiencies, and it is plotted in Fig. 5b for 780 nm as a function of the decay lengths shown in Fig. 5a. The vertical error bars are based on 2 standard deviations of 10 measurements, and the horizontal error bars are the vertical error bars from Fig. 5a. Although the curves shown here are a product of efficiencies, they should have the same shape as the input efficiency (recall that the output gratings are the same for all grating pairs). This confirms that there is a peak in input efficiency for a certain groove scattering strength, as predicted earlier. Moreover, this result shows that the coupling efficiency for an improperly chosen groove size can easily be 10 times smaller than that of an optimal groove size. Note that the best coupling is produced at rather shallow groove depths (between 35 nm and 70 nm); deeper grooves scatter out too strongly, not allowing the SPP power to build up.

According to Fig. 5b, the peak efficiency occurs for gratings with SPP decay length of ≈ 12 μm. Using the theoretically derived relationship $L_s = L_a$ at the peak (i.e. radiative loss equals nonradiative loss), we calculate a nonradiative power decay length (i.e. $L_a/2$) of ≈ 24 μm. This value is less than the nonradiative decay length in the Au film (60 μm), and it shows that the presence of grooves in the film increases the absorption loss. The added loss is most likely caused by the roughness and other fabrication defects in the grooves. The loss also seems to depend on the aspect ratio of the grooves. For example, in Fig. 5b the peak for the wider, shallower grooves is slightly higher and to the right of the peak for narrower, deeper grooves. This suggests that the wider grooves are less lossy and result in higher coupling efficiencies. However, the effect of aspect ratio on efficiency is small compared to the effect of scattering strength. In other words, we get a large gain in efficiency by choosing a decay length corresponding to a peak in one of the curves in Fig. 5b and only a small additional gain by choosing the curve with the highest peak.

The input and output efficiencies cannot be extracted separately from the product plotted in Fig. 5b. Nevertheless, lower and upper bounds can be placed on them to estimate the individual coupler performance. Note that the output grating groove size happens to be the same as the input grating groove size for one of the peaks in Fig. 5b (width 110 nm, depth 47 nm). This implies that $L_s = L_a$ for all the output gratings (all output gratings are the same), in which case the total radiative scattering at the output equals the nonradiative loss. Therefore, the output efficiency is at most 50 % (it could be less than that if, for example, some of the outscattered light falls outside the NA of the objective). By using this result together with the data in Fig. 5b (peak input-output efficiency of ≈ 22.5 %), it can be calculated that the peak input coupling efficiency at 780 nm exceeds 45 %. This means that at least 45 % of the light incident on the input grating is converted into SPPs.

Finally, we measured the intensity profiles of the input gratings to check for consistency with the theoretical predictions (Eq. 8, Fig. 2). The results for the 190 nm wide grooves at 780 nm wavelength are shown in Fig. 6. The error bars here represent two standard deviations of 10 different measurements. We compare them to theoretical profiles obtained using Eq. 8 with $L_s / L_a$ and $A_0$ being adjustable parameters, while the parameter $\frac{1}{2}(1/L_s + 1/L_a)^{-1}$ is set equal to the measured SPP power decay lengths (Fig. 5a); in addition, a common uniform vertical offset is added to account for the background intensity in the image. These results confirm that there is a transition between the two regimes illustrated earlier in Fig. 2. Groove depth 6 in Fig. 6 exhibits the $L_s < L_a$ behavior, and depth 7 exhibits the $L_s > L_a$ behavior, suggesting that the maximum input efficiency should be between depths 6 and 7. Indeed, this is consistent with the measured input-output efficiency product in Fig. 5b, in which the peak occurs at depth 6. Similar agreements were observed for gratings with 110 nm and 60 nm widths (not shown here for the sake of brevity). This confirms that measured intensity profiles in the direction perpendicular to the grooves can be used to estimate if a grating is too strongly or too weakly scattering for optimal coupling, i.e. if it is to the right or left of the maximum efficiency peak in Fig. 5b.
In addition, while the precise measurement of $L_s$ and $L_a$ for each grating was not the main focus of the paper, the combination of the measured total decay length $\frac{1}{2}(1/L_s + 1/L_a)^{-1}$ reported in Fig. 5a and the values of the ratio $L_s/L_a$ obtained from the fits to Fig. 6 can in principle be used to determine the two parameters separately.

Three key observations can be made from the results presented above. First, the grating coupler efficiency is a single-peaked function of scattering strength, which depends on groove dimensions. Second, the groove aspect ratio has a small effect on efficiency; therefore, efficiency can be optimized by fixing one dimension (width or depth) and varying the other. Third, the measured input intensity profile is an indicator of whether the scattering strength should be increased or decreased to improve efficiency. Based on these three findings, one can easily optimize a grating coupler by testing several devices with varying groove width or depth.

The same types of measurements were performed at 675 nm, leading to qualitatively similar results (presented as electronic supplementary material). The peak efficiency at that wavelength is 30 % compared to 45 % at 780 nm and occurs at shorter SPP decay lengths (9 μm instead of 12 μm). This is expected since the absorption loss in Au increases at the shorter wavelength, resulting in shorter $L_a$ and therefore shorter $L_s$ at the peak. To summarize, the highest-efficiency coupler that we demonstrated at 780 nm had groove pitch, width, and depth of 560 nm, 190 nm, and 38 nm respectively; the highest-efficiency coupler at 675 nm had the same groove pitch and width, while the groove depth was 28 nm. Both couplers were 30 μm by 40 μm in size.

# Conclusion

We have presented the design and testing of a large-area grating coupler that coverts light into SPPs with efficiencies exceeding 45 % at 780 nm and 30 % at 675 nm. The coupler has a classical geometry of rectangular grooves on an Au film that is simple to fabricate. The optimization was performed by choosing a grating period such that there are no multiple diffraction orders in air, and by tuning the size of the grooves to strike a balance between out-scattering of SPPs and in-coupling of light. The optimal groove dimensions were determined experimentally by using the input grating intensity profile as an indicator of groove scattering strength. Furthermore, we showed that the grating efficiency can be more than an order of magnitude lower if the groove dimensions are not properly chosen. We believe that this work is applicable to a variety of plasmonic devices that require efficient coupling of light into SPPs.

# Acknowledgments

This work is supported by the NIST-CNST NanoFab, NIST Division 637 FIB-SEM, and the NIST-CNST/UMD-Nanocenter Cooperative Agreement. We also thank Prof. Igor Griva for useful discussions.

## Figures

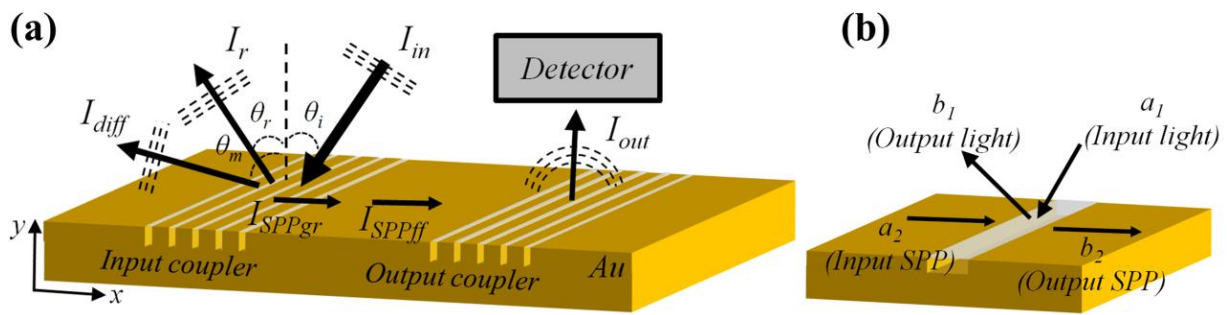

**Fig. 1** a) Gratings converting free-space light into SPP (input coupler) and vice versa (output coupler). $I_{in}$, $I_r$, $I_{diff}$, $I_{SPPgr}$, $I_{SPPff}$ and $I_{out}$ are the intensities of the input beam, reflected beam, diffracted beam, grating SPP, flat film SPP, and out-coupled beam, respectively; $\theta_i$, $\theta_r$, and $\theta_m$ are the incidence, reflection and diffraction angles, respectively. b) Electromagnetic modes scattered from a single groove at the input, schematically shown by their wave vectors and amplitudes.

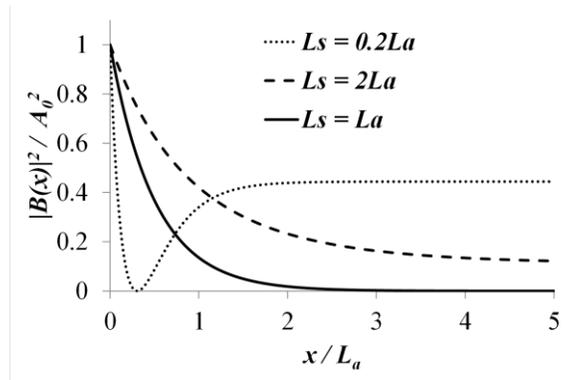

**Fig. 2** Normalized theoretical intensity of light reflected by the input grating as a function of position on the grating for different values of $L_s$ (scattering decay length) relative to $L_a$ (absorption decay length).

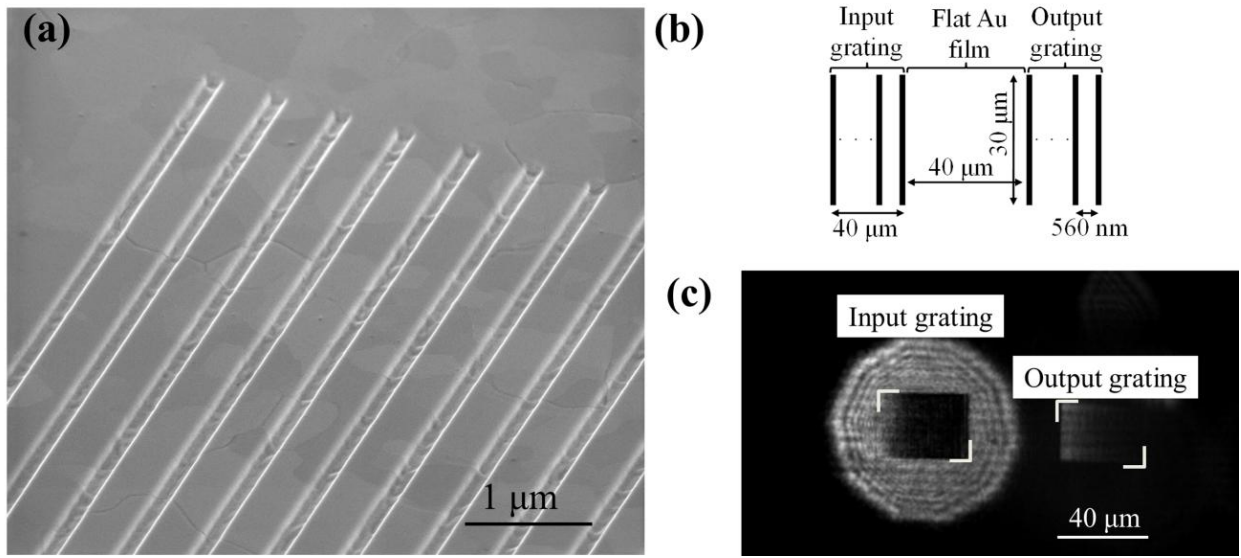

**Fig. 3** a) Scanning electron micrograph of fabricated grating coupler. The Au film on mica is atomically smooth except at the grain boundaries . b) Layout of input-output grating coupler pairs. In most couplers, the spacing between gratings is 40 μm; in coupler pairs used to estimate the SPP propagation loss in the flat Au film, this spacing is varied. c) Microscope image of a grating pair under laser illumination of the input grating at a wavelength of 780 nm. The incidence angle at the input grating (left) is adjusted such than an SPP is excited and travels to the output grating (right), which then scatters it out. The corner marks outline the grating boundaries. The bright circular region around the input grating is the specular reflection from the illuminated areas of flat Au film.

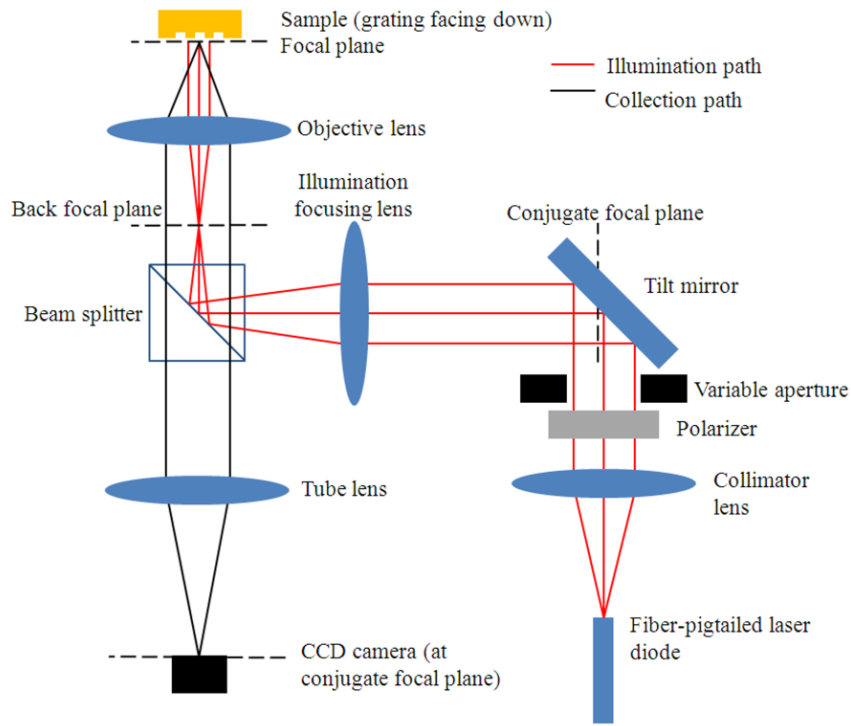

**Fig. 4** Schematic of experimental setup used to test grating couplers.

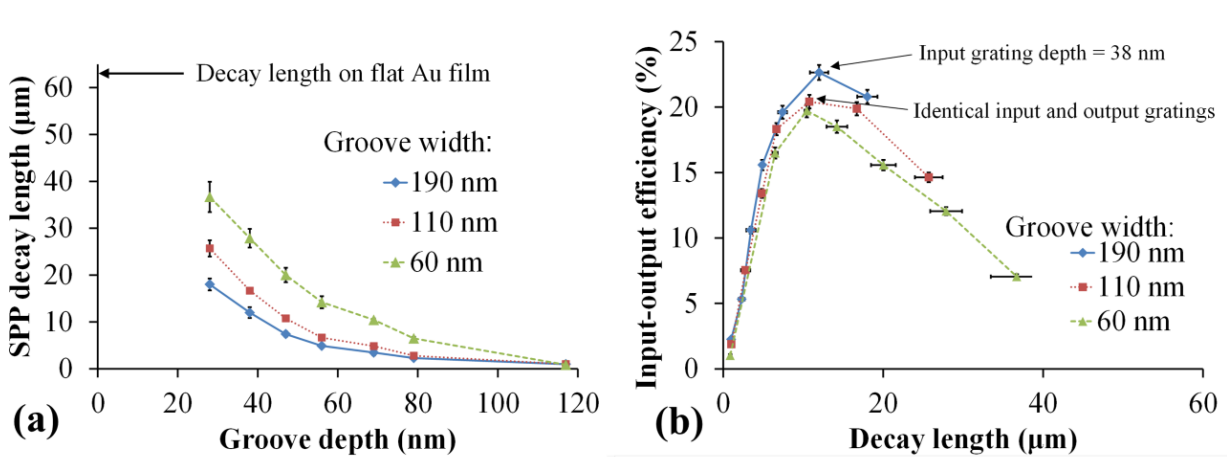

**Fig. 5** a) Measured SPP power decay length along grating surfaces as a function of groove depth, plotted for three different groove widths (780 nm wavelength). b) Product of input and output grating coupling efficiencies as a function of SPP decay length on the input grating, plotted for three different input grating groove widths (780 nm wavelength). Decay lengths are related experimentally to groove depth as shown in a).

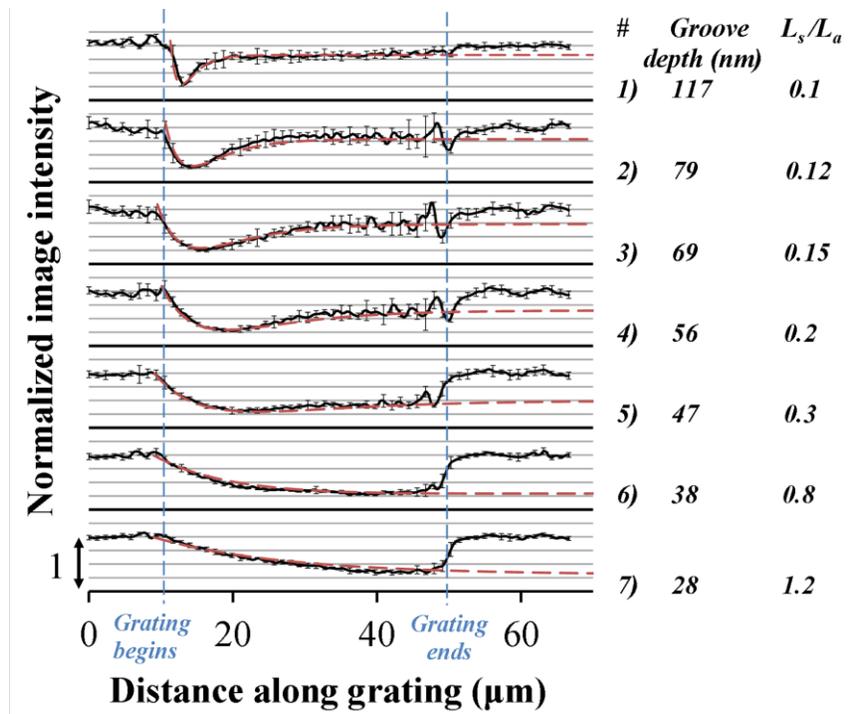

**Fig. 6** Solid curves: measured intensity profiles extracted from images of input gratings with varying groove depths (groove width 190 nm, wavelength 780 nm). Dashed curves: theoretical profiles based on Eq. 8; the $L_s / L_a$ ratio for each profile is listed. The thick horizontal lines represent zero intensity for each curve. The oscillations near the end of the grating are standing waves caused by small SPP reflection at the grating edge.

## Electronic Supplementary Material

File ESM1.pdf contains the following supplementary material:
1) Fig. S1 showing SPP decay length on gratings at 675 nm 2) Fig. S2 showing the product of input and output grating efficiencies at 675 nm 3) Fig. S3 showing input grating intensity profiles at 675 nm.

# An Efficient Large-Area Grating Coupler for Surface Plasmon Polaritons

*Stephan T. Koev[1,2], Amit Agrawal[1,2,3], Henri J. Lezec[1], and Vladimir Aksyuk[1,*]*

[1]*Center for Nanoscale Science and Technology, National Institute of Standards and Technology, Gaithersburg, MD 20899, USA*

[2]*Maryland Nanocenter, University of Maryland, College Park, MD 20742, USA*

[3]*Department of Electrical Engineering and Computer Science, Syracuse University, Syracuse, NY 13244, USA*

*Corresponding Author: vaksyuk@nist.gov

## Electronic Supplementary Materials

The experimental results presented below are similar to those in Figures 5 and 6 in the text, but they are for measurements at 675 instead of 780 nm wavelength. The samples used, experimental procedures, and error bar conventions are the same.

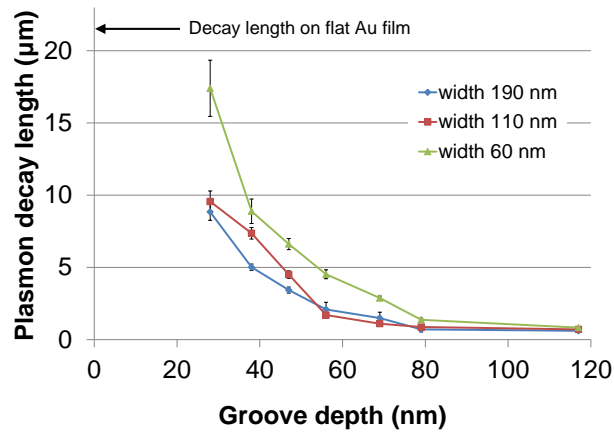

**Fig. S1** Measured SPP decay length on grating surfaces as a function of groove depth, plotted for three different groove widths (675 nm wavelength).

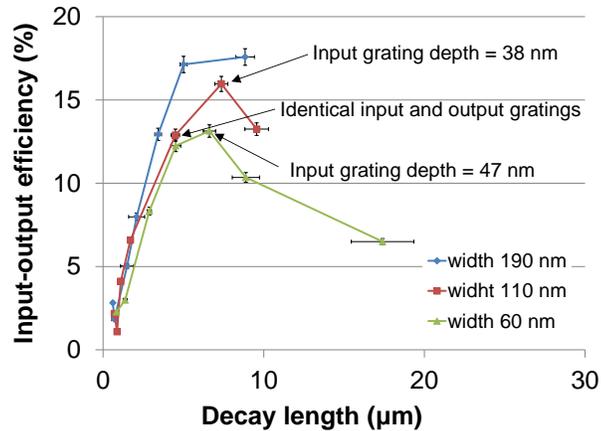

**Fig. S2** Product of input and output grating coupling efficiencies as a function of SPP decay length on the input grating, plotted for tree different input grating groove widths (675 nm wavelength). We estimated that the output coupler efficiency* is less than 58 %; therefore, the input coupler efficiency corresponding to the peak in the figure exceeds 30 %.

*The upper bound on the output efficiency is estimated based on testing the coupler pairs in reverse, i.e. using the side with variably sized grooves as the output. When the grating with groove width 110 nm and groove depth 38 nm is used as an input (forward measurement), it corresponds to a peak in efficiency (Figure S2). As explained in the text, an input grating is maximally efficient when $L_s = L_a$, which means that its output efficiency can be at most 50 % (half the light is absorbed and half is scattered out). When that grating is used as an output (reverse measurement), the input-output efficiency product is 11.7 ± 0.3 % (uncertainty based on two standard deviations of 10 measurements). Therefore, the input efficiency of the "standard" grating exceeds 22.7 % (the standard grating is the one that is the same in all coupler pairs; its groove width and depth are 110 nm and 47 nm respectively). The measured input-output efficiency product of the symmetric coupler pair (both sides have the "standard" grating) is 12.9 ± 0.4 %. From this, we calculate that the output efficiency of the standard coupler is less than 58 %.

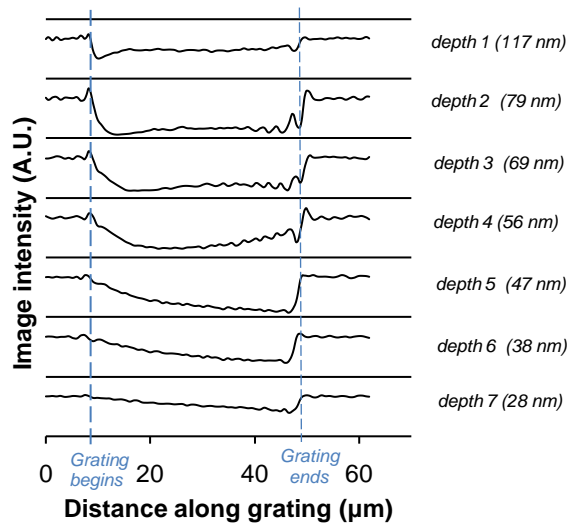

**Fig. S3** Intensity profiles extracted from images of input gratings with varying groove depths (groove width 60 nm, wavelength 675 nm). The thick horizontal lines represent zero intensity for each curve. Depth 4 exhibits $L_s < L_a$ behavior, and depth 5 exhibits the $L_s > L_a$ behavior, suggesting that the maximum efficiency should be between depths 4 and 5. This agrees with the measured efficiency in Figure S2, in which the peak occurs at depth 5 for the 60 nm groove width.